\documentclass[12pt,prd,showpacs,nofootinbib]{revtex4}

\usepackage{amssymb,epsfig,longtable}

\begin{document}
\title[BTZ solution with torsion]{Exact vacuum solution of a
  $(1+2)$--dimensional Poincar\'e gauge theory: BTZ solution with
  torsion}
\author{Alberto A.\ Garc\'{\i}a}
\email{aagarcia@fis.cinvestav.mx}
\affiliation{Departamento de F\'{\i}sica, CINVESTAV--IPN\\
Apartado Postal 14--740, C.P. 07000, M\'exico, D.F., Mexico}
\author{Friedrich W.\ Hehl}
\email{hehl@thp.uni-koeln.de} \affiliation{Institute for
Theoretical Physics, University of Cologne, 50923 K\"oln,
Germany\\
{\rm and} Department of Physics and Astronomy\\
University of Missouri-Columbia, Columbia, MO 65211, USA}
\author{Christian Heinicke}\email{
chh@thp.uni-koeln.de} \affiliation{Institute for Theoretical
Physics, University of Cologne, 50923 K\"oln, Germany}
\author{Alfredo Mac\'{\i}as}
\email{amac@xanum.uam.mx} \affiliation{Departamento de
F\'{\i}sica, Universidad Aut\'onoma Metropolitana--Iztapalapa\\
Apartado Postal 55--534, C.P. 09340, M\'exico, D.F., Mexico}

\begin{abstract}
  In the framework of (1+2)--dimensional Poincar\'e gauge gravity, we
  start from the Lagrangian of the Mielke--Baekler (MB) model that
  depends on torsion and curvature and includes {\em translational\/}
  and {\em Lorentzian\/} Chern--Simons terms. We find a general
  stationary circularly symmetric vacuum solution of the field
  equations. We determine the properties of this solution, in
  particular its mass and its angular momentum. For vanishing torsion,
  we recover the BTZ--solution. We also derive the general conformally
  flat vacuum solution with torsion. In this framework, we discuss
  {\em Cartan's} (3--dimensional) {\em spiral staircase} and find that
  it is not only a special case of our new vacuum solution, but can
  alternatively be understood as a solution of the 3--dimensional
  Einstein--Cartan theory with matter of constant pressure and
  constant torque. {\em file 3dexact19.tex, 2003-06-21}
\end{abstract}

\pacs{04.20.Jb, 04.90.+e, 11.15.-q}
\keywords{Einstein-Cartan theory, Chern-Simons terms,
topological gravity, torsion}
\maketitle

\section{Introduction}

On first sight, $(1+2)$--dimensional gravity seems to be rather
boring. In 3 dimensions (3D), the Weyl tensor vanishes and the
curvature is fully determined by the Ricci tensor and thus, via
the Einstein equation, by the energy-momentum alone.  Outside the
sources the curvature is zero and there are no propagating degrees
of freedom, i.e., no gravitational waves. Moreover, there is no
Newtonian limit. But even if spacetime is flat, it is not trivial
globally. A point particle, e.g., generates the spacetime geometry
of a cone. In such a geometry we have light bending, double
images, etc. The spacetimes for N particles can be constructed
similarly by gluing together patches of $(1+2)$D Minkowski
space.  This was occasionally investigated since the late 1950s,
see Deser et al. \cite{deser84} and the review of Carlip
\cite{Carlip98}.

Some problems in $(1+3)$D gravity reduce to an effective ($1+2$)D
theory, like the cosmic string, e.g.; the high--temperature behavior
of ($1+3$)D theories also motivates the study of $(1+2)$D theories. In
this context, Deser, Jackiw, and Tempelton (DJT) proposed a $(1+2)$D
gravitational gauge model with topologically generated mass
\cite{deser82}. However, the real push for $(1+2)$D gravitational
models came when Witten formulated the $(1+2)$D Einstein theory as a
Chern--Simons theory, in a similar way as proposed by Ach\'ucarro and
Townsend \cite{amat}, and showed its exact solvability in terms of a
finite number of degrees of freedom \cite{witten88,witten89}. Also de
Sitter gravity, conformal gravity, and supergravity, in $(1+2)$D, turn
out to be equivalent to Chern--Simons theories
\cite{horne89,koehler90,dereli00,DereliS}, see also the recent
monograph of Blagojevi\'c \cite{blagojevic}.

Mielke and Baekler (MB) proposed a $(1+2)$D topological gauge model
with torsion and curvature \cite{mielke91,baekler92} from which the
DJT--model can be derived by imposing the constraint of vanishing
torsion by means of a Lagrange multiplier term. Gravitational theories
in $(1+2)$D with torsion, see also Tresguerres \cite{tresguerres92}
and Kawai \cite{kawai94}, are analogous to the continuum theory of
lattice defects in crystal physics, in particular, the corresponding
theory of dislocations relates to a torsion of the underlying
continuum, see Kr\"oner \cite{kroener81}, Kleinert \cite{kleinert},
Dereli and Ver\c{c}in \cite{dereli87,dereli91}, Katanaev and Volovich
\cite{katanaev91}, and Kohler \cite{kohler95}.  The fresh approach of
Lazar \cite{lazar00,lazar02_1,lazar02_2} promises additional insight.

The next important impact on $(1+2)$D gravity was the discovery of
a black hole solution by Ba\~{n}ados, Teitelboim, and Zanelli
(BTZ) \cite{banados92}. The BTZ black hole is locally isometric to
anti-de~Sitter (AdS) spacetime. It can be obtained, see Brill
\cite{brill}, from the AdS spacetime as a quotient of the latter
with the group of finite isometries. It is asymptotically anti--de
Sitter and has no curvature singularity at the origin.
Nevertheless, it is clearly a black hole: it has an event horizon
and, in the rotating case, an inner horizon. Also electrically and
magnetically charged generalizations are known. For extensive
discussions see the reviews
\cite{banados93,carlip95_1,carlip95_2,Carlip98,banados99,birmingham01}.
The relevance to $(1+3)$D gravity can also be seen from the fact
that the BTZ solution can be derived from the $(1+3)$D
Pleba\'nski--Carter metric by means of a dimensional reduction
procedure, see Cataldo et al.\ \cite {cataldo00}. By means
of the BTZ solution, many interesting questions can be addressed
in the context of quantum gravity. For example, Strominger
computed the entropy of the BTZ black hole microscopically
\cite{strominger98}. There is also a relationship between the BTZ
black hole and string theory, see Hemming and Keski--Vakkuri
\cite{hemming01}.

Thus, although $(1+2)$D gravity lacks many important features of real,
$(1+3)$D gravity, it keeps enough characteristic structure to be of
interest, especially in view of the fact that in the $(1+2)$D case
many calculations can be done which are far too involved in $(1+3)$D
for the time being.

In this paper we show that the BTZ-metric can be embedded in the
framework of the specific Poincar\'e gauge model proposed by
Mielke and Baekler. We arrive at a ``BTZ-solution with torsion'',
see Table 1, and discuss some of its characteristic properties.---

In section II, we introduce briefly the MB--model and its field
equations. In vacuum, these yield constant torsion and constant
curvature and, by a suitable ansatz, we obtain the new solution
displayed in Table 1. In section III, we discuss some of the
properties of our new solution.  In particular, we compute its
quasi--local energy and angular momentum expressions as it was
suggested to us by Nester, Chen, Tung, and Wu
\cite{chen94,chen99,chen00,nester00,wu01,wu03}.

In section IV we derive the general conformally flat vacuum solution
and show its relation to the solution of Table 1. In the final section
V, we point out that Cartan's spiral staircase, an example of a
simple non--Euclidean connection that is constructed {}from 3D
Euclidean space, can be understood as a specific vacuum solution of
the MB-model as well as a solution of 3D Einstein--Cartan theory with
matter of constant pressure and constant torque.


\section{Mielke-Baekler model and its BTZ-like exact solution}

Our geometric arena is 3D Riemann-Cartan space.  The basic variables
are the {\em coframe} $\vartheta^\alpha = e_i{}^\alpha \, dx^i$ and
the {\em Lorentz connection} $\Gamma^{\alpha\beta} =
\Gamma_i{}^{\alpha\beta} \, dx^i $. Latin letters $i,j,\,\dots=0,1,2$
denote holonomic or coordinate indices and Greek letters
$\alpha,\beta,\,\dots=\hat 0,\hat 1,\hat 2$ anholonomic or frame
indices. In an {\em orthonormal coframe}, which we assume for the rest
of our article, the metric is given by $g=-\vartheta^{\hat 0} \otimes
\vartheta^{\hat 0} + \vartheta^{\hat 1} \otimes \vartheta^{\hat 1} +
\vartheta^{\hat 2} \otimes \vartheta^{\hat 2}$. In such an orthonormal
coframe, the connection is antisymmetric $\Gamma^{\alpha\beta}=
-\Gamma^{\beta\alpha}$. The frame dual to the coframe reads
$e_\alpha=e^i{}_\alpha\,\partial_i$, with $e_\alpha \rfloor
\vartheta^\beta=\delta_\alpha^\beta$, where $\rfloor$ denotes the
interior product.  We introduce the abbreviation
$\vartheta^{\alpha\beta \dots}:= \vartheta^\alpha\wedge
\vartheta^\beta\wedge\cdots$ and the $\eta$-basis (${}^\star$ denotes
the Hodge-star operator) $\eta := {}^\star1 \,, \quad \eta_{\alpha} :=
{}^\star\vartheta_\alpha \,, \quad \eta_{\alpha\beta} :=
{}^\star\vartheta_{\alpha\beta} \,,\quad \eta_{\alpha\beta\gamma} :=
{}^\star\vartheta_{\alpha\beta\gamma} \,.  $ In 3 dimensions,
$\eta_{\alpha\beta\gamma}$ is the totally antisymmetric unit tensor.
For our conventions, one should compare \cite{hehl95}.

{}From the gauge potentials coframe and connection, we can derive the
field strengths {\em torsion} and {\em curvature} ($D$ denotes the exterior
covariant derivative),
\begin{equation}\label{structure}
  T^\alpha :=D
  \vartheta^\alpha = d\vartheta^\alpha +\Gamma_\beta{}^\alpha \wedge
  \vartheta^\beta \,, \quad
  R_\alpha{}^\beta
:= d\Gamma_\alpha{}^\beta - \Gamma_\alpha{}^\gamma
  \wedge \Gamma_\gamma{}^\beta\,.
\end{equation}

In a Riemann-Cartan space, the connection can be expressed in terms of the
torsion and the anholonomity 2-form $\Omega^\alpha := d \vartheta^\alpha$,
\begin{equation}\label{Gamma}
  \Gamma_{\alpha\beta} = e_{[\alpha} \rfloor \Omega_{\beta]} - \frac{1}{2}
  \, \left( e_\alpha \rfloor e_\beta \rfloor \Omega_\gamma \right) \,
  \vartheta^\gamma \nonumber - e_{[\alpha}\rfloor T_{\beta]} +
  \frac{1}{2} \, \left( e_\alpha \rfloor e_\beta \rfloor T_\gamma
  \right) \vartheta^\gamma\,,
\end{equation}
see \cite{hehl95} Eq.(3.10.6), for $dg_{\alpha\beta}=0$ and
$Q_{\alpha\beta}=0$.

Mielke and Baekler \cite{mielke91,baekler92} considered the following
Lagrangian:
\begin{eqnarray}\label{mblag}
  V_{\rm MB}
&=&
-\frac{\chi}{2\ell} \, R^{\alpha\beta} \wedge \eta_{\alpha\beta}
-\frac{\Lambda}{\ell} \, \eta
+\frac{\theta_{{\rm T}}}{2\ell^2} \, \vartheta^\alpha \wedge T_\alpha
\nonumber \\
&&-\frac{\theta_{{\rm L}}}{2} \, \left(\Gamma_\alpha{}^\beta
    \wedge d \Gamma_\beta{}^\alpha - \frac{2}{3} \,
    \Gamma_\alpha{}^\beta \wedge \Gamma_\beta{}^\gamma \wedge
    \Gamma_\gamma{}^\alpha \right)
+ L_{\rm mat}\,.
\end{eqnarray}
The first term, the usual Einstein-Cartan term, is followed by the
cosmological term and the Chern-Simons terms for torsion and
curvature, see \cite{hehl91}. The last term denotes the matter
Lagrangian that is minimally coupled to gravity.  The 3D gravitational
constant $\ell$ guarantees dimensional consistency.  The
Einstein-Cartan piece is multiplied by a dimensionless constant
$\chi$, with $\chi=1$ or $\chi=0$, and the Chern-Simons pieces by the
dimensionless ``vacuum angles'' $\theta_{\rm T}$ and $\theta_{\rm L}$.

{}From this model we can derive the Deser-Jackiw-Tempelton (DJT) model
of topological massive gravity \cite{deser82} by adding a Lagrange
multiplier term $\lambda_\alpha\,T^\alpha$ to the Lagrangian $V_{\rm
  MB}$ thereby enforcing {\em vanishing torsion\/}. Quite recently,
Blagojevi\'c and Vasili\'c \cite{blagojevic03} considered a restricted
MB-model with $\theta^2_{\rm T}+\chi\Lambda\ell^2=0$, $\,\theta_{\rm
  L}=0$, and $\chi=1$, which yields, in vacuum, the field equation
$R^{\alpha\beta}=0$, i.e., {\em vanishing curvature}, introducing
thereby the teleparallel geometry of empty spacetime dynamically. A
similar teleparallel model (including torsion square terms) was
developed by Sousa and Maluf \cite{maluf03}.

We find the field equations by variation of (\ref{mblag}) with respect
to coframe and (Lorentz) connection:
\begin{eqnarray}
\label{field1}
\frac{\chi}{2} \, \eta_{\alpha\beta\gamma} \, R^{\beta\gamma} +
\Lambda \, \eta_{\alpha} - \frac{\theta_{{\rm T}}}{\ell} \,
T_\alpha &=& \ell \, \Sigma_\alpha\,,\\
\label{field2} \frac{\chi}{2} \, \eta_{\alpha\beta\gamma} \, T^{\gamma} -
\frac{\theta_{{\rm T}}}{2\ell} \, \vartheta_{\alpha\beta} -
\theta_{{\rm L}} \, \ell \, R_{\alpha\beta} &=& \ell \,
\tau_{\alpha\beta} \,.
\end{eqnarray}
The 2-forms of the material energy-momentum and spin currents are
defined by $\Sigma_\alpha:=\delta L_{\rm mat}/\delta \theta^\alpha$
and $\tau_{\alpha\beta}:=\delta L_{\rm mat}/\delta\Gamma_\alpha
{}^\beta$, respectively.

The field equations represent inhomogeneous algebraic equations in
torsion $T^\alpha$ and curvature $R^{\alpha\beta}$. We can resolve
them with respect to $T^\alpha$ and $R^{\alpha\beta}$
\cite{mielke91,baekler92}.  The {\em vacuum} field equations result by
equating $\Sigma_\alpha$ and $\tau_{\alpha\beta}$ to zero.  Then, by
assuming $\chi^2+2\theta_{{\rm T}}\theta_{{\rm L}}\neq 0$, we obtain
$T_\alpha =2 {\cal T} \eta_\alpha /\ell$ and $R_{\alpha\beta} = {\cal
  R} \vartheta_{\alpha\beta} / \ell^2$; for the definitions of ${\cal
  T}$ and ${\cal R}$, see Table 1. The torsion has only an {\em axial}
part and, similarly, the curvature a {\em scalar} part, both with 1
independent component.

A solution is specified by a pair $(\vartheta^\alpha,
\Gamma^{\alpha\beta})$.  We start with a static and circularly
symmetric (orthonormal) coframe,
\begin{equation}\label{eq6}
\vartheta^{\hat t}  = N(r) \, dt \,,\quad
\vartheta^{\hat r} =  \frac{\displaystyle dr}{\displaystyle N(r)} \,, \quad
\vartheta^{\hat \phi}  =  G(r) \, \left[-W(r) \, dt + d\phi\right] \,,
\end{equation}
where $N(r),\,G(r)$, and $W(r)$ are free functions. Since the torsion
is known from the field equations, we can substitute it, together with
(\ref{eq6}), into (\ref{Gamma}). This yields $\Gamma_{\alpha\beta}$
which, together with the known curvature, leads to
\begin{eqnarray}
G=A+Br\,,\qquad W=\frac{\alpha}{(A+br)^2}+\beta \,,\\
N^2(r)=C+\frac{\alpha^2}{(rB+A)^2} -\frac{\Lambda_{{\rm
        eff}}}{B^2} \, \left(A^2-2ABr-B^2r^2\right)\,,
\end{eqnarray}
where $A,B,C,\alpha,\beta$ are integration constants. Moreover, we
introduced an effective cosmological constant $\Lambda_{\rm eff}$, see
Table 1. By means of the coordinate transformation $r\to Ar+B$ and
$\phi\to\phi+\beta \, t$ and some change in notation, we arrive at our
new BTZ-like solution with torsion, see Table 1 for its explicit form.
The topological terms in the Lagrangian will induce an effective
cosmological constant even if the `bare' cosmological constant
$\Lambda$ vanishes.  If we put ${\theta}_{{\rm L}}={\theta}_{{\rm
    T}}=0$, then $\Lambda_{\rm eff}=-\Lambda$ and $T^\alpha=0$, and we
fall back to the standard BTZ solution \cite{banados92}.

\begin{table*}
\caption{Exact vacuum solution of the 3D
  Poincar\'e gauge model of Mielke--Baekler: BTZ--like solution with
  torsion}
\begin{tabular}{|l|l|}
\hline
\begin{tabular}{l}
vacuum\\
field equations
\end{tabular}
&
\parbox{9.5cm}{
\begin{eqnarray*}
  \frac{\chi}{2} \, \eta_{\alpha\beta\gamma} \, R^{\beta\gamma} +
  \Lambda \, \eta_{\alpha} - \frac{\theta_{{\rm T}}}{\ell} \,
  T_\alpha &=& 0\\ \frac{\chi}{2} \, \eta_{\alpha\beta\gamma} \,
  T^{\gamma} - \frac{\theta_{{\rm T}}}{2\ell} \,
  \vartheta_{\alpha\beta} - \theta_{{\rm L}} \,
  \ell \, R_{\alpha\beta}& = & 0\,
\end{eqnarray*}}\\
\hline \hline coframe & \parbox{9.5cm}{
\begin{eqnarray*}
  \vartheta^{\hat t} & = & \psi(r) \, dt \\ \vartheta^{\hat r} & = &
  \frac{ dr}{ \psi(r)} \qquad \qquad {\scriptsize \psi(r) :=
    \sqrt{\left(\frac{J}{2r}\right)^2 - M + \Lambda_{\rm eff} \,r^2}}
  \\ \vartheta^{\hat \phi} & = & r\, \left(-\frac{ J}{ 2 r^2}\,dt+
    d\phi\right)
\end{eqnarray*}}\\
\hline metric & \parbox{9.5cm}{
\begin{displaymath}
g= - \vartheta^{\hat t} \otimes \vartheta^{\hat t}
   + \vartheta^{\hat r} \otimes \vartheta^{\hat r}
   + \vartheta^{\hat \phi} \otimes \vartheta^{\hat \phi}
\end{displaymath}}\\
\hline connection &
\parbox{9.5cm}{
\begin{eqnarray*}
  \Gamma^{\hat t \hat r} = -\Gamma^{\hat r \hat t} &=&
  \left(\frac{{\cal T}}{\ell} \, \frac{J}{2r} - \Lambda_{\rm eff}
    \,r\right) dt + \left(\frac{J}{2r} - \frac{{\cal T}}{\ell} \,
    r\right) \, d\phi \\ \Gamma^{\hat r \hat \phi} = - \Gamma^{\hat
    \phi \hat r} &=& \psi(r) \, \left(\frac{{\cal T}}{\ell} \, dt +
    d\phi\right) \\ \Gamma^{\hat \phi \hat t} = - \Gamma^{\hat t \hat
    \phi} &=& -\left( \frac{J}{2r^2} + \frac{{\cal T}}{\ell}\right) \,
  \frac{dr}{\psi(r)}
\end{eqnarray*}}\\
\hline torsion &
\parbox{9.5cm}{
\begin{displaymath}
T^{\alpha} = 2\,\,\frac{{{\cal T}}}{\ell} \, \eta^{\alpha}
\end{displaymath}}\\
\hline curvature &
\begin{tabular}{ll}
Riemann-Cartan &
\parbox{5cm}{
\begin{displaymath}
R^{\alpha\beta} = \frac{{\cal R}}{\ell^2} \,
\vartheta^{\alpha\beta}
\end{displaymath}}\\
Riemann &
\parbox{5cm}{
\begin{displaymath}
  \widetilde{R}^{\alpha\beta} = \Lambda_{\rm eff} \,
  \vartheta^{\alpha\beta}
\end{displaymath}}
\end{tabular}\\
\hline Cotton &
\begin{tabular}{ll}
Riemann-Cartan &
\parbox{5cm}{
\begin{displaymath}
  C^{\alpha} = -\frac{{\cal T}\,{\cal R}}{\ell^3} \, \eta^{\alpha}
\end{displaymath}}\\
Riemann &\parbox{5cm}{\begin{displaymath} \widetilde
    C^{\alpha}=0
\end{displaymath}}
\end{tabular}\\
\hline\hline
\begin{tabular}{l}
constants\\
\end{tabular}
&
\parbox{9.5cm}{\begin{displaymath}
{{\cal T}} := \frac{-\frac{\theta_{{\rm T}}}{2}\, \chi +\Lambda
\ell^2 \theta_{{\rm L}}} {\chi^2+2\theta_{{\rm T}} \theta_{{\rm
L}}} \qquad {\cal R}:=-\frac{\theta_{{\rm T}}^2+\chi \Lambda
\ell^2}{\chi^2+2\theta_{{\rm T}} \theta_{{\rm L}}}
\end{displaymath}
\begin{displaymath}\hspace{0pt}
\Lambda_{\rm eff}:= \frac{{\cal T}^2+{\cal R}}{\ell^2}
\end{displaymath}}\\
\hline
\end{tabular}
\end{table*}
\normalsize

\section{Properties of our solution}

\subsection{Autoparallels and extremals}

In a Riemann--Cartan space, the autoparallels (straightest lines)
and the extremals or geodesics (longest/shortest lines) do not
coincide in general. An autoparallel curve $x^i(s)$ obeys, in
terms of a suitable affine parameter $s$, the equation
\begin{equation}\label{auto}
  \frac{d^2 \, x^k(s)}{ds^2} + \Gamma_{ij}{}^k \, \frac{d \,
    x^i(s)}{ds} \, \frac{d \, x^j(s)}{ds} = 0 \,.
\end{equation}
The (holonomic) components of the connection $\Gamma_{ij}{}^k$
depend on metric and torsion according to
\begin{equation}\label{decomp}
  \Gamma_{ij}{}^k = \widetilde{\Gamma}_{ij}{}^k - \, K_{ij}{}^k\,,
  \qquad K_{ij}{}^k := \frac{1}{2}\left(-T_{ij}{}^k
    +T_j{}^k{}_i-T^k{}_{ij}\right)\,,
\end{equation}
where $\widetilde{\Gamma}_{ij}{}^k$ is the Christoffel symbol and
$K_{ij}{}^k$ the contortion. In (\ref{auto}), only the symmetric
part of the connection enters. By means of (\ref{decomp}), it can
be expressed as
\begin{equation}\label{verweis}
  \Gamma_{(ij)}{}^k= \widetilde{\Gamma}_{(ij)}{}^k + T^k{}_{(ij)}\,.
\end{equation}
The extremals are determined by the metrical properties of spacetime
alone and follow from the variation of the world length $\int
\sqrt{-g_{ij} \, \dot x^i \, \dot x^j}$ in the standard way:
\begin{equation}
  \frac{d^2 \, x^k(s)}{ds^2} + \widetilde{\Gamma}_{ij}{}^k \, \frac{d
    \, x^i(s)}{d s}\frac{d \, x^j(s)}{ds} = 0 \,.
\end{equation}
For our solution, see Table 1,
\begin{equation}
  T_{ijk} = 2 \, \frac{\cal T}{\ell} \, \eta_{ijk}
  \qquad\Longrightarrow\qquad T_{i(jk)}=0\,.
\end{equation}
Thus, the torsion dependent piece drops out in (\ref{verweis}) and
(\ref{auto}).  Autoparallels and extremals coincide and we get the
same geodesics as in the case of the standard BTZ--solution in
Riemannian spacetime.

\subsection{Killing vectors}

In a Riemann-Cartan space we call $\xi = \xi^\alpha \, e_{\alpha}$ a
Killing vector if the latter is the generator of a symmetry
transformation of the metric and of the connection according to
\begin{equation}\label{killing1}
  {\pounds}_\xi \, g =0 \,,\quad \pounds_\xi \, \Gamma_\alpha{}^\beta
  = 0 \,,
\end{equation}
see \cite[p.83]{hehl95}.
These two relations can be recast into a more convenient form,
\begin{eqnarray}
  e_{(\alpha} \rfloor \widetilde D\xi_{\beta)} &=& 0 \,,\\
  D\left(e_\alpha \rfloor \stackrel{\frown}{D} \xi^\beta\right) +\xi
  \rfloor R_\alpha{}^\beta &=&0\,,
\end{eqnarray}
where $\widetilde D$ refers to the Riemannian part of the
connection (Levi--Civita connection) and $\stackrel{\frown}{D}$ to
the transposed connection: $\stackrel{\frown}{D}:=d +
\stackrel{\frown}{\Gamma}_\alpha{}^\beta :=d+
\Gamma_\alpha{}^\beta + e_\alpha \rfloor T^\beta$. For our
solution we find two Killing vectors, namely
\begin{equation}
  \stackrel{(t)}{\xi} := \partial_t \quad \mbox{and} \quad
  \stackrel{(\phi)}{\xi} := \partial_\phi \,,
\end{equation}
that is, the same Killing vectors as in the case of the standard
BTZ solution.

\subsection{Quasilocal conserved quantities}

Now we consider the conserved quantities of our solution. Nester,
Chen, and Wu \cite{nester00}, see also the literature quoted there,
proposed a quasi--local boundary expression within metric--affine
gravity, a theory the spacetime of which goes beyond the
Riemann--Cartan structure in that it carries additionally a
nonmetricity.  We adapt the formulas of \cite{nester00} for the case
of vanishing nonmetricty. The derivation starts from a first--order
Lagrange $n$--form $V$ that is at most quadratic in its field
strengths $T^\alpha$ and $R^{\alpha\beta}$. The corresponding momenta
read $ H_\alpha := -{\partial \, V}/{\partial T^\alpha}$ and $
H_{\alpha\beta} := -{\partial \, V}/{\partial R^{\alpha\beta}}$.  The
Lagrangian can be decomposed with respect to a vector field $N$, with
$N\rfloor d\nu=1$:
\begin{eqnarray}
V &= &d\nu \wedge N \rfloor V \nonumber \\
&=:& d\nu \wedge \Big[ - \left(\pounds_N \vartheta^\alpha \right)
\wedge H_\alpha -\left(\pounds_N \Gamma_\alpha{}^\beta \right)
\wedge H^\alpha{}_\beta - N^\alpha{\mathfrak H_\alpha} -
d\,{\mathfrak B} \Big]\,.
\end{eqnarray}
The Hamilton $2$--form $\mathfrak H$ is defined by ${\mathfrak
  H}:=N^\alpha \, {\mathfrak H}_{\alpha} + d\,{\mathfrak B}$ . Since
$\mathfrak H_\alpha$ turns out to be proportional to the field
equations, only the spatial boundary $1$--form $\mathfrak B$
contributes to the boundary integral of $\mathfrak H$. In order to
obtain finite values for the quasi--local ``charges'', the boundary
term has to be compared to a reference or background solution which
will be denoted by a bar over the corresponding symbol. As background,
we choose our solution with $M=0,\,J=0$. Moreover, the difference of a
quantity $\alpha$ between a solution and the background is
$\Delta\alpha := \alpha - \overline{\alpha}$.  Then, the quasi--local
charges are given by \cite{nester00}
\begin{eqnarray}
{\mathfrak B}(N) &:=& - \left\{
\begin{array}{rcl}
  (N\rfloor \vartheta^\alpha ) \, \Delta H_\alpha &+& \Delta
  \vartheta^\alpha \left(N \rfloor \overline{H}_\alpha\right)\\
  (N \rfloor \overline{\vartheta}^\alpha) \, \Delta H_\alpha &+&
  \Delta \vartheta^\alpha \left(N \rfloor H_\alpha \right)
\end{array}\right\} \nonumber \\
&& - \left\{\begin{array}{rcl} (\stackrel{\frown}{D}{}^\alpha
N^\beta )
    \, \Delta H_{\alpha\beta} &+& \Delta \Gamma^{\alpha\beta}
    \left(N \rfloor \overline{H}_{\alpha\beta} \right) \vspace{5pt}\\
    \overline{(\stackrel{\frown}{D}{}^\alpha N^\beta)} \, \Delta
    H_{\alpha\beta} &+& \Delta\Gamma^{\alpha\beta} \left(N
      \rfloor H_{\alpha\beta}\right)
\end{array}\right\}\,.
\end{eqnarray}
The upper (lower) line in the braces is chosen if the field strengths
(momenta) are prescribed on the boundary.  The momenta of our solution
read $H_\alpha = -({\theta_{{\rm T}}}/{2\ell^2}) \, \vartheta_\alpha$
and $H_{\alpha\beta} = (\chi /2\ell) \, \eta_{\alpha\beta} -
(\theta_{{\rm L}}/2) \, \Gamma_{\alpha\beta}$.

We derive the quasi--local energy and angular momentum by taking
for the vector field $N$ the Killing vectors $\partial_t$ or
$\partial_\phi$, respectively:
\begin{eqnarray}
  \ell\, {\mathfrak B}(\partial_t) &=& \left[\theta_{{\rm L}}
    \left(\Lambda_{\rm eff}\ell\,J -{\cal T}\, M\right) + {\chi}
    \left(\Lambda_{\rm eff} r^2 -\sqrt{\Lambda_{\rm eff}} \,r\psi
    \right) \right] d \phi \nonumber \\ && -\frac{1}{2\ell} \, \left[
    (2\theta_{\rm L} \, {\cal T}^2 -\theta_{\rm T}) \, M -
    2\ell\theta_{\rm L} \, \Lambda_{\rm eff} J {\cal T} +\chi \, (\ell
    \Lambda_{\rm eff} J - 2M{\cal T})\right] \, dt,\\ \ell\,
  {\mathfrak B}(\partial_\phi) &=& - \left[\frac{\chi}{2} \, J +
    \theta_{{\rm L}} \left(\ell\,M- {\cal T}\,J\right)\right] d\phi
  \nonumber \\ && - \left[ {\chi} \left( \Lambda_{\rm eff} r^2
      -\sqrt{\Lambda_{\rm eff}} \, r\psi \right) +\frac{1}{\ell}
    \left( {\theta_{{\rm L}}} {\cal T} - {\chi} \right) \left(\ell\, M
      -{\cal T} \, J \right) + \frac{\theta_{\rm T}}{2\ell}\, J
  \right] dt\,.
\end{eqnarray}
We assume the existence of the Einstein-Cartan piece, i.e., $\chi=1$.
In order to obtain total energy and angular momentum, we have to
integrate, for $t={\rm const}$, the ${\mathfrak B}$'s over a full
circle and to perform the limit $r\to \infty$.  For ${\cal
  T}=\theta_{\rm T}=\theta_{\rm L}=0$, our solution reduces to the
standard BTZ one. In that case, total energy and total angular
momentum reduce to $M$ and $J$.  Thus, in our conventions, the
gravitational constant is $\ell = \pi$.  Moreover, as in general
relativity, see Wald \cite[p.296]{wald84}, we introduce a factor
$-1$ into the angular momentum:
\begin{eqnarray}
  E_\infty &=& \frac{1}{\pi}\,\lim_{r\to\infty} \, \int_0^{2\pi}
  \left[ \theta_{{\rm L}} \left(\Lambda_{\rm eff}\ell\,J - {\cal
        T}\,M\right) +  \left(\Lambda_{\rm eff} r^2
      -\sqrt{\Lambda_{\rm eff}} \,r \psi \right) \right] d \phi
  \nonumber \\ &=& M - 2\theta_{{\rm L}}\left( {\cal T}\,M -
  \Lambda_{\rm eff}\ell\,J\right) \,,\\ L_{\infty} & = &
  (-1)\frac{1}{\pi}\,\lim_{r\to\infty} \, \int_0^{2\pi} -
  \left[\frac{1}{2} \, J + \theta_{{\rm L}} \left(\ell\,M- {\cal
        T}\,J\right)\right] d\phi \nonumber \\ &=& J + 2\theta_{{\rm L}}
  \left(\ell\, M - {\cal T}\,J\right) \,.
\end{eqnarray}
Thus, for $\theta_{{\rm L}}=0$, the two integration constants $M$
and $J$ have their conventional interpretation as energy (mass)
and angular momentum, as with the BTZ--metric in general
relativity. However, for $\theta_{{\rm L}}\ne 0$, we find in each
case admixtures from the other ``charge'', respectively. This is
not too surprising, since torsion and curvature emerge in both
field equations.


\section{General conformally flat vacuum solution with torsion}

The vacuum field equations of the MB model imply constant {\em
  Riemann--Cartan} curvature and constant Riemannian curvature. The
Cotton 2--form reads
\begin{equation}
  C_{\alpha} := DL_\alpha \,, \qquad L_\alpha := e_\beta \rfloor
  R_\alpha{}^\beta + \frac{1}{2(n-1)} \left( e_\beta\rfloor \,
    e_\gamma\rfloor R^{\beta\gamma}\right)\vartheta_\alpha \,.
\end{equation}
The Riemannian Cotton 2--form is zero. Thus, the metric is conformally
flat, see, e.g., \cite{garcia03}.  Hence the ansatz
\begin{equation}
\vartheta^{\hat 0} = \frac{dt}{\Psi} \,, \qquad \vartheta^{\hat 1}
= \frac{dx}{\Psi} \,, \qquad \vartheta^{\hat 2} = \frac{dy}{\Psi}
\,,
\end{equation}
where $\Psi=\Psi(t,x,y)$, via the 1st field equation, yields,
\begin{equation}
\Psi = \Psi^{(t)}(t) + \Psi^{(x)}(x) + \Psi^{(y)}(y) \,,
\end{equation}
\begin{equation}
-\partial_{xx} \, \Psi^{(x)} = \partial_{tt} \, \Psi^{(t)} =
-\partial_{yy} \Psi^{(y)}\,.
\end{equation}
This leads to a general solution with 5 parameters $A,B,C,D,E$,
\begin{equation}
\Psi = A \, \left(-t^2+x^2+y^2\right) + Bt+Cx+Dy+E\,,
\end{equation}
with one constraint on the parameters,
\begin{equation}
0=B^2-C^2-D^2+6AE + \Lambda_{\rm eff} \,.
\end{equation}
For $B=C=D=0$, $E=1$ we recover the usual form of the (anti--)de
Sitter metric, for $A=B=D=E=0$ the Poincar\'e metric. Coordinate
transformations that yield the BTZ--metric are given in
\cite{Carlip98}.

In the anti--de Sitter case, the solution reads
\begin{equation}\label{eq59}
  \vartheta^\alpha = \frac{dx^\alpha}{\psi} \,,\qquad
  \psi=1-\frac{\Lambda_{\rm eff}}{6} (-t^2+x^2+y^2) \,,
\end{equation}
\begin{equation}\label{eq60}
  \Gamma^{\alpha\beta} = \frac{{\cal T}}{\ell} \, \eta^{\alpha\beta} +
  x^{[\alpha} \, \vartheta^{\beta]} \, \frac{\Lambda_{\rm eff}}{3} \,.
\end{equation}
For $\theta_T=0$, we recover the solution of Dereli and Ver\c{c}in
\cite{dereli91}.

If the coupling constants are chosen such that
\begin{equation}\label{telecons}
\theta_{\rm T}^2+\chi\Lambda\ell^2 = 0\,,
\end{equation}
the Riemann--Cartan curvature is zero $R_{\alpha\beta}=0$ and the
torsion reduces to $ {\cal T} = \ell \, \sqrt{\Lambda_{\rm eff}}$. We
obtain a {\em teleparallel subcase} of the MB--model. The teleparallel
sector of the MB--model, defined by (\ref{telecons}) and $\theta_{\rm
  L}=0$, is extensively studied in \cite{blagojevic03}, see also the
closely related cases \cite{maluf03,park98,fjelstad01}.  We stress
that our exact solution carries both, torsion {\em and\/} curvature.
Therefore it is more general and should be carefully distinguished
from its teleparallel limit.

\section{\'E.\ Cartan's spiral staircase}

If we put $\Lambda_{\rm eff}=0$, then, see (\ref{eq59}) and
(\ref{eq60}), we arrive at
\begin{equation}\label{stair1}
  \vartheta^\alpha=\delta_i^\alpha\,dx^i\,,\qquad
  \Gamma^{\alpha\beta}=\frac{\cal T}{\ell}\,\eta^{\alpha\beta}\,.
\end{equation}
The components of the connection are totally antisymmetric:
$\Gamma_{\gamma\alpha\beta}=e_\gamma\rfloor\Gamma_{\alpha\beta}=({\cal
  T}/\ell)\,\eta_{\gamma\alpha\beta}$. The Riemannian curvature
vanishes. By simple algebra we find,
\begin{equation}\label{stair2}
  T^{\alpha} = 2\,\,\frac{{{\cal T}}}{\ell} \, \eta^{\alpha}\,,
  \qquad\widetilde{R}^{\alpha\beta} =0\,,\qquad R^{\alpha\beta} =-
  \frac{{\cal T}^2}{\ell^2} \, \vartheta^{\alpha\beta}\,.
\end{equation}
This is a subcase of our solution of Table 1.

In fact, for Euclidean signature, we recover Cartan's {spiral 3D
  staircase\/} of 1922 \cite{cartan22}, see Fig.\ 1:
\begin{quote}
  ``\dots imagine a space F which corresponds point by point with a
  Euclidean space E, the correspondence preserving distances. The
  difference between the two space is following: two orthogonal triads
  issuing from two points A and A' infinitesimally nearby in F will be
  parallel when the corresponding triads in E may be deduced one from
  the other by a given helicoidal displacement (of right--handed sense,
  for example), having as its axis the line joining the origins. The
  straight lines in F thus correspond to the straight lines in E: They
  are geodesics. The space F thus defined admits a six parameter group
  of transformations; it would be our ordinary space as viewed by
  observers whose perceptions have been twisted.  Mechanically, it
  corresponds to a medium having constant pressure and constant internal
  torque.''
\end{quote}

\begin{figure}
\caption{
  {\em Cartan's spiral staircase.\/} Cartan's rules \cite{cartan22}
  for the introduction of a non-Euclidean connection in a 3D Euclidean
  space are as follows: (i) A vector which is parallelly transported
  along itself does not change (cf.\ a vector directed and transported
  in $x$-direction). (ii) A vector that is orthogonal to the direction
  of transport rotates with a prescribed constant `velocity'' (cf.\ a
  vector in $y$--direction transported in $x$--direction).  The
  winding sense around the three coordinate axes is always positive.}
\begin{center}
\epsfig{file=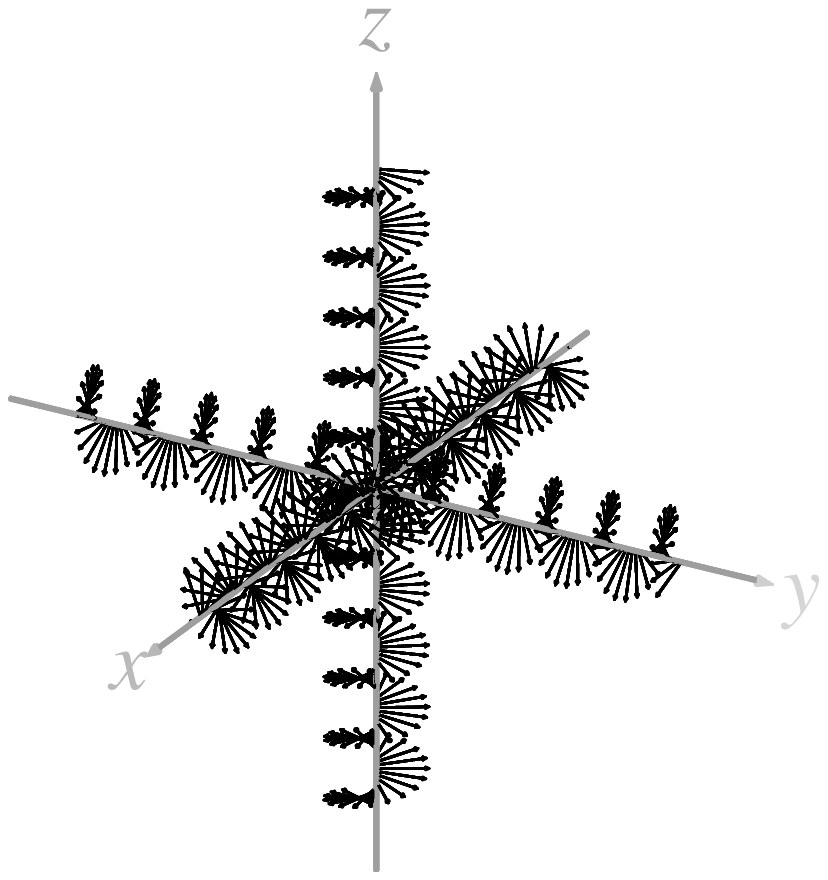,width=8cm}
\end{center}
\end{figure}

Obviously, Cartan's prescriptions are reflected in the solution
(\ref{stair1}). For (\ref{stair1}), autoparallels and extremals
coincide. Thus, in the spiral staircase, extremals are {\em
  Euclidean} straight lines. This is apparent in Cartan's
construction.

Cartan apparently had in mind a 3D space with Euclidean signature.
For an alternative interpretation of Cartan's spiral staircase we
consider the 3D Einstein--Cartan field equations without cosmological
constant:
\begin{eqnarray}
  \frac{1}{2} \, \eta_{\alpha\beta\gamma} \, R^{\beta\gamma}
&=&\ell \, \Sigma_\alpha \,,\\ \frac{1}{2} \, \eta_{
    \alpha\beta\gamma} \, T^\gamma &=& \ell \, \tau_{\alpha\beta} \,.
\end{eqnarray}
The coframe and the connection of (\ref{stair1}), Euclidean signature
assumed, form a solution of the Einstein--Cartan field equations {\em
  with matter} provided the energy--momentum current (for Euclidean
signature the force stress tensor ${\mathfrak t}_\alpha{}^\beta$) and
the spin current (here the torque or moment stress tensor ${\mathfrak
  s}_{\alpha\beta}{}^\gamma$) are constant,
\begin{equation}
  \Sigma_{\alpha} =:{\mathfrak t}_\alpha{}^\beta\,\eta_\beta = -
  \frac{{\cal T}^2}{\ell^3}
\,\eta_\alpha\,\quad{\rm and}\quad
 \tau_{\alpha\beta} =:{\mathfrak s}_{\alpha\beta}{}^\gamma\,\eta_\gamma =
 - \frac{\cal T}{\ell^2} \,
  \vartheta_{\alpha\beta}\,.
\end{equation}
Inversion yields
\begin{equation}
  {\mathfrak t}_\alpha{}^\beta= -\frac{{\cal
      T}^2}{\ell^3} \, 
\delta_\alpha^\beta\,,\qquad {\mathfrak
    s}_{\alpha\beta\gamma}= -\frac{\cal T}{\ell^2} \,\eta_{\alpha \beta
    \gamma}\,.
\end{equation}
We find a constant hydrostatic {\em pressure} $ -{\cal T}^2/\ell^3$
and a constant {\em torque} $-{\cal T}/\ell^2$, exactly as foreseen by
Cartan.  In solid state physics, this corresponds to a superposition
of three ``forests'' of {\em screw dislocations} that are parallel to
the coordinate axes with constant and equal densities. However, in a
real crystal, the Riemann--Cartan curvature $R^{\alpha\beta}$ has to
vanish (instead of the Riemannian curvature
$\widetilde{R}^{\alpha\beta}$, as in our exact solution) and no
pressure would emerge macroscopically.

Thus we can either view the spiral staircase as a vacuum solution
and special case of our solution of Table 1 or as a material
solution of 3D Einstein--Cartan theory (with Euclidean signature)
carrying constant pressure and constant torque.

\vspace{-0.3truecm}

\begin{acknowledgments}
  We thank Yuri Obukhov (Moscow) for a critical reading of our paper
  and for many suggestions. Helpful remarks of Milutin Blagojevi\'c
  (Belgrade) are also greatly appreciated.  One of the authors (fwh)
  is grateful to Jim Nester, Chiang-Mei Chen, Roh-Suan Tung, and
  Yu-Huei Wu (all of Chung-Li) for discussions on quasilocal energy.
  This Mexican/German work has been supported by the CONACYT Grants
  42191--F, and 38495--E, by the CONACYT--DFG Grant E130--655, and by
  the DFG Grant 444 MEX--113/12/6-1.
\end{acknowledgments}

\end{document}